\documentclass[12pt,a4paper]{article}
\usepackage{amsmath}
\usepackage{bm}
\usepackage{amsfonts}
\usepackage{graphicx}
\usepackage[normal]{caption2}
\usepackage{subfigure}
\usepackage{rotating}
\usepackage{citesort}
\usepackage{lscape}
\usepackage{section}
\begin{document}
\renewcommand\arraystretch{1.1}
\setlength{\abovecaptionskip}{0.1cm}
\setlength{\belowcaptionskip}{0.5cm}
\pagestyle{empty}
\newpage
\pagestyle{plain} \setcounter{page}{1} \setcounter{lofdepth}{2}
\begin{center} {\large\bf Confrontation of QMD model with the experimental data for $^{40}Ar$ +$^{45}$Sc, $^{197}Au$ +$^{197}$Au and $^{129}Xe$
+$^{119}$Sn reactions}\\
\vspace*{0.4cm}
{\bf Tajinder Pal Singh$^a$} and {\bf Sakshi Gautam$^b$} \footnote{Email:~sakshigautm@gmail.com} \\
$^a${\it  Handloom street, H.No.-B-9-1076,Surgapuri,Kotkapura
(Distt. Faridkot)-151204, India\\} $^b${\it  Department of
Physics, Panjab University, Chandigarh -160 014, India.\\}
\end{center}
In the present work, we make confrontation of our theoretical
calculations using quantum molecular dynamics model with the
experimental data for the reactions of $^{40}Ar$ +$^{45}$Sc,
$^{197}Au$ +$^{197}$Au and $^{129}Xe$ +$^{119}$Sn at different
incident energies. In these reactions, we display the charge
distribution and energy dependence of fragments multiplicity. Our
results indicate good agreement with the experimental data for all
the reactions .
\newpage
\baselineskip 20pt
\section{Introduction}
 \par
Nuclear physics, in general and heavy-ion collisions, in
particular, are of central interest due to several rare phenomena
emerging at different incident energies. These include (1)
collective flow and its disappearance [1], (2) breaking of
colliding nuclei into pieces i.e. multi-fragmentation [2,3], (3)
sub-threshold particle production [4] and (4) formation of hot and
dense nuclear matter [5]. At high excitation energy, the colliding
nuclei may break-up into several small and intermediate size
fragments along with large number of nucleons that are also
emitted. This is known as multi-fragmentation. The
multi-fragmentation is the complex phenomena in nature which
depends crucially on the incident energy of the projectile and on
the geometry of the reaction [6]. At very low excitation energies,
the excitation energy deposited in the system is too small to
allow the break up of the nuclei into fragments. With the increase
in the incident energy, the nuclei, after collision, can break up
into dozens of fragments consisting of light, medium and heavy
fragments. This phenomena was observed with the help of
accelerators by Perfilov \emph{et al.} and Lozkin \emph{et al.}
[7,8]. After 1982, Jakobsson \emph{et al.} [9] observed a multiple
emission of fragments of medium mass known as intermediate mass
fragments (IMF's) in emulsion irradiation by the carbon beam of
250 MeV/nucleon at Berkeley BEVALAC. This scenario of
multi-fragmentation was discussed in 1983 by Siemens [10].
Actually, the nuclear multi-fragmentation was discovered through
cosmic rays accompanying the collisions of relativistic protons
with targets and following the emission of slow fragments. These
fragments were heavier than alpha particles but lighter than
fission fragments. Also, the rise and fall of IMFs has also been
established by Ogilvie \emph{et al.} [11] through the collisions
of Au particles with C, Al and Cu targets using ALADIN forward
spectrometer at GSI, Darmstadt, with beam accelerated by SIS
synchrotron.
\par
The development of accelerators led to huge experimental data on
multifragmentation. On theoretical front also, a large number of
models have been proposed to study multifragmentation. These
include statistical models and dynamical models. In the present
paper, we aim to show the validity of quantum molecular dynamics
model by confronting the calculations with the experimental data
over wide range of energies.
 \par
\section{The Formalism}
\subsection{Quantum Molecular Dynamics (QMD) model}
\par
We describe the time evolution of a heavy-ion reaction within the
framework of Quantum Molecular Dynamics (QMD) model [12-15] which
is based on a molecular dynamics picture. Here each nucleon is
represented by a coherent state of the form
\begin{equation}
\phi_{\alpha}(x_1,t)=\left({\frac {2}{L \pi}}\right)^{\frac
{3}{4}} e^{-(x_1-x_{\alpha }(t))^2}
e^{ip_{\alpha}(x_1-x_{\alpha})} e^{-\frac {i p_{\alpha}^2 t}{2m}}.
\label {e1}
\end{equation}
where L is the Gaussian width of the particle. Thus, the wave
function has two time dependent parameters $x_{\alpha}$ and
$p_{\alpha}$. The total n-body wave function is assumed to be a
direct product of coherent states:
\begin{equation}
\phi=\phi_{\alpha}
(x_1,x_{\alpha},p_{\alpha},t)\phi_{\beta}(x_2,x_{\beta},
p_{\beta},t)....,         \label {e2}
\end{equation}
where antisymmetrization is neglected. One should, however, keep
in the mind that the Pauli principle, which is very important at
low incident energies, has been taken into account. The initial
values of the parameters are chosen in a way that the ensemble
($A_T$+$A_P$) nucleons give a proper density distribution as well
as a proper momentum distribution of the projectile and target
nuclei. The time evolution of the system is calculated using the
generalized variational principle. We start out from the action
\begin{equation}
S=\int_{t_1}^{t_2} {\cal {L}} [\phi,\phi^{*}] d\tau, \label {e3}
\end{equation}
with the Lagrange functional
\begin{equation}
{\cal {L}} =\left(\phi\left|i\hbar \frac
{d}{dt}-H\right|\phi\right), \label {e4}
\end{equation}
where the total time derivative includes the derivatives with
respect to the parameters. The time evolution is obtained by the
requirement that the action is stationary under the allowed
variation of the wave function
\begin{equation}
\delta S=\delta \int_{t_1}^{t_2} {\cal {L}} [\phi ,\phi^{*}] dt=0.
\label{e5}
\end{equation}
If the true solution of the Schr\"odinger equation is contained in
the restricted set of wave function
$\phi_{\alpha}\left({x_{1},x_{\alpha},p_{\alpha}}\right),$ this
variation of the action gives the exact solution of the
Schr\"odinger equation. If the parameter space is too restricted,
we obtain that wave function in the restricted parameter space
which comes close to the solution of the Schr\"odinger equation.
Performing the variation with the test wave function (2), we
obtain for each parameter $\lambda$ an Euler-Lagrange equation;
\begin{equation}
\frac{d}{dt} \frac{\partial {\cal {L}}}{\partial {\dot
{\lambda}}}-\frac{\partial \cal {L}} {\partial \lambda}=0.
\label{e6}
\end{equation}
For each coherent state and a Hamiltonian of the form, \\

$H=\sum_{\alpha}
\left[T_{\alpha}+{\frac{1}{2}}\sum_{\alpha\beta}V_{\alpha\beta}\right]$,
the Lagrangian and the Euler-Lagrange function can be easily
calculated \cite{aich91}
\begin{equation}
{\cal {L}} = \sum_{\alpha}{\dot {\bf x}_{\alpha}} {\bf
p}_{\alpha}-\sum_{\beta} \langle{V_{\alpha
\beta}}\rangle-\frac{3}{2Lm}, \label{e7}
\end{equation}
\begin{equation}
{\dot {\bf x}_{\alpha}}=\frac{{\bf
p}_\alpha}{m}+\nabla_{p_{\alpha}}\sum_{\beta} \langle{V_{\alpha
\beta}}\rangle, \label {e8}
\end{equation}
\begin{equation}
{\dot {\bf p}_{\alpha}}=-\nabla_{{\bf x}_{\alpha}}\sum_{\beta}
\langle{V_{\alpha \beta}}\rangle. \label {e9}
\end{equation}
Thus, the variational approach has reduced the n-body
Schr\"odinger equation to a set of 6n-different equations for the
parameters which can be solved numerically. If one inspects  the
formalism carefully, one finds that the interaction potential
which is actually the Br\"{u}ckner G-matrix can be divided into
two parts: (i) a real part and (ii) an imaginary part. The real
part of the potential acts like a potential whereas imaginary part
is proportional to the cross section.

In the present model, interaction potential comprises of the
following terms:
\begin{equation}
V_{\alpha\beta} = V_{loc}^{2} + V_{loc}^{3} + V_{Coul} + V_{Yuk},
\label {e10}
\end{equation}
$V_{loc}$ is the Skyrme force whereas $V_{Coul}$ and $V_{Yuk}$
define, respectively, the Coulomb and Yukawa potentials. Yukawa
term separates surface which also play role in low energy process
like fusion and cluster radioactivity [16,17]. The expectation
value of these potentials is calculated as
\begin{eqnarray}
V^2_{loc}& =& \int f_{\alpha} ({\bf p}_{\alpha}, {\bf r}_{\alpha},
t) f_{\beta}({\bf p}_{\beta}, {\bf r}_{\beta}, t)V_I ^{(2)}({\bf
r}_{\alpha}, {\bf r}_{\beta})
\nonumber\\
&  & \times {d^{3} {\bf r}_{\alpha} d^{3} {\bf r}_{\beta}
d^{3}{\bf p}_{\alpha}  d^{3}{\bf p}_{\beta},}
\end{eqnarray}
\begin{eqnarray}
V^3_{loc}& =& \int  f_{\alpha} ({\bf p}_{\alpha}, {\bf
r}_{\alpha}, t) f_{\beta}({\bf p}_{\beta}, {\bf r}_{\beta},t)
f_{\gamma} ({\bf p}_{\gamma}, {\bf r}_{\gamma}, t)
\nonumber\\
&  & \times  V_I^{(3)} ({\bf r}_{\alpha},{\bf r}_{\beta},{\bf
r}_{\gamma}) d^{3} {\bf r}_{\alpha} d^{3} {\bf r}_{\beta} d^{3}
{\bf r}_{\gamma}
\nonumber\\
&  & \times d^{3} {\bf p}_{\alpha}d^{3} {\bf p}_{\beta} d^{3} {\bf
p}_{\gamma}.
\end{eqnarray}
where $f_{\alpha}({\bf p}_{\alpha}, {\bf r}_{\alpha}, t)$ is the
Wigner density which corresponds to the wave functions (eq. 2). If
we deal with the local Skyrme force only, we get
\small{\begin{equation} V^{Skyrme} = \sum_{{\alpha}=1}^{A_T+A_P}
\left[\frac {A}{2} \sum_{{\beta}=1} \left(\frac
{\tilde{\rho}_{\alpha \beta}}{\rho_0}\right) + \frac
{B}{C+1}\sum_{{\beta}\ne {\alpha}} \left(\frac {\tilde
{\rho}_{\alpha \beta}} {\rho_0}\right)^C\right].
\end{equation}}

 Here A, B and C are the Skyrme parameters which are
defined according to the ground state properties of a nucleus.
Different values of C lead to different equations of state. A
larger value of C (= 380 MeV) is often dubbed as stiff equation of
state.The finite range Yukawa ($V_{Yuk}$) and effective Coulomb
potential ($V_{Coul}$) read as:
\begin{equation}
V_{Yuk} = \sum_{j, i\neq j} t_{3}
\frac{exp\{-|\textbf{r}_{\textbf{i}}-\textbf{r}_{\textbf{j}}|\}/\mu}{|\textbf{r}_{\textbf{i}}-\textbf{r}_{\textbf{j}}|/\mu},
\end{equation}
\begin{equation}
V_{Coul} = \sum_{j, i\neq
j}\frac{Z_{eff}^{2}e^{2}}{|\textbf{r}_{\textbf{i}}-\textbf{r}_{\textbf{j}}|}.
\end{equation}
\par
The Yukawa interaction (with $t_{3}$= -6.66 MeV and $\mu$ = 1.5
fm) is essential for the surface effects. Also, the relativistic
effect does not play role in low incident energy of present
interest [18].
\par
The phase space of the nucleons is stored at several time steps.
The QMD model does not give any information about the fragments
observed at the final stage of the reaction. In order to construct
fragments from the present phase-space, one needs the
clusterization algorithms. So, to construct fragments, we use
Minimum Spanning Tree (MST) method.
\par\begin{figure}[!t]
\centering
 \vskip 1cm
\includegraphics[angle=0,width=12cm]{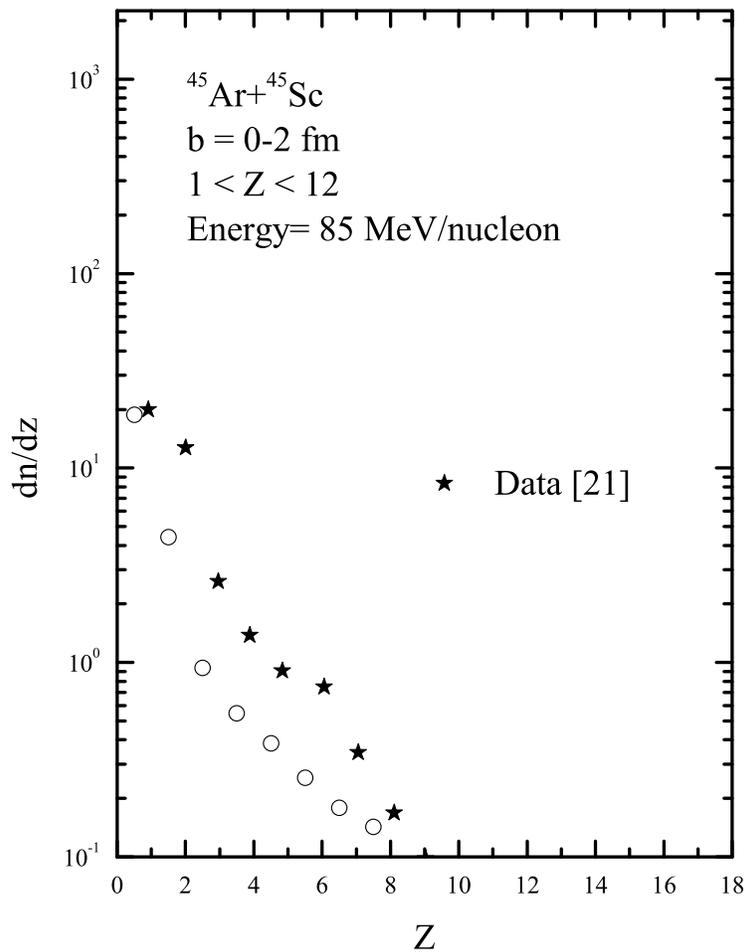}
 \vskip -0cm \caption{ The charge distribution (dn/dz) for the central collision of
  $^{40}Ar$ +$^{45}$Sc at an incident energy of 85 MeV/nucleon. The experimental data has been taken by the Ref. [21].}\label{fig1}
\end{figure}

\section{Minimum spanning tree (MST) method}
 The widely used clusterization algorithm is the Minimum
Spanning Tree (MST) method [19]. In MST method, two nucleons are
allowed to share the same fragment if their centroids are closer
than a distance $r_{min}$,
\begin{equation}
|\textbf{r}_{\textbf{i}}-\textbf{r}_{\textbf{j}}| \leq r_{min}.
\end{equation}
where $\textbf{r}_{\textbf{i}}$ and $\textbf{r}_{\textbf{j}}$ are
the spatial positions of both nucleons. The value of $r_{min}$ can
vary between 2-4 fm. This method gives a big fragment at high
density which splits into several light and medium mass fragments
after several hundred fm/c. This procedure gives same fragment
pattern for times later than 200 fm/c, but cannot be used for
earlier times.

\begin{figure}[!t]
\centering
 \vskip 1cm
\includegraphics[angle=0,width=12cm]{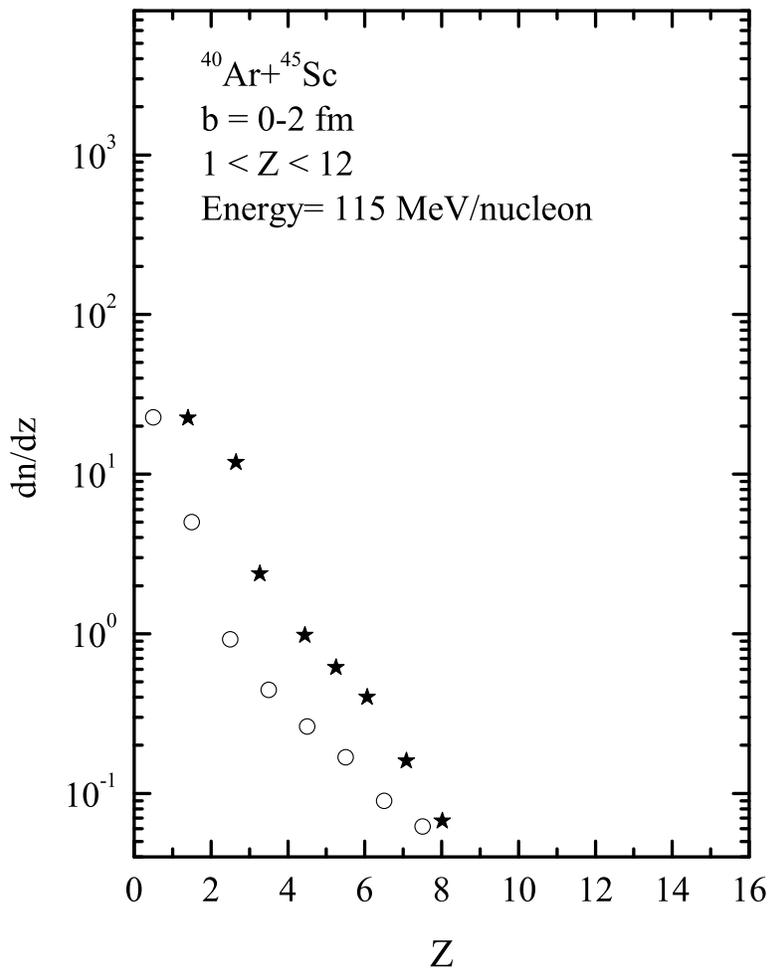}
 \vskip -0cm \caption{Same as figure 1, but at an incident energy of 115 MeV/nucleon. The experimental data has been taken by the Ref. [21].}\label{fig2}
\end{figure}

\begin{figure}[!t]
\centering
 \vskip 1cm
\includegraphics[angle=0,width=12cm]{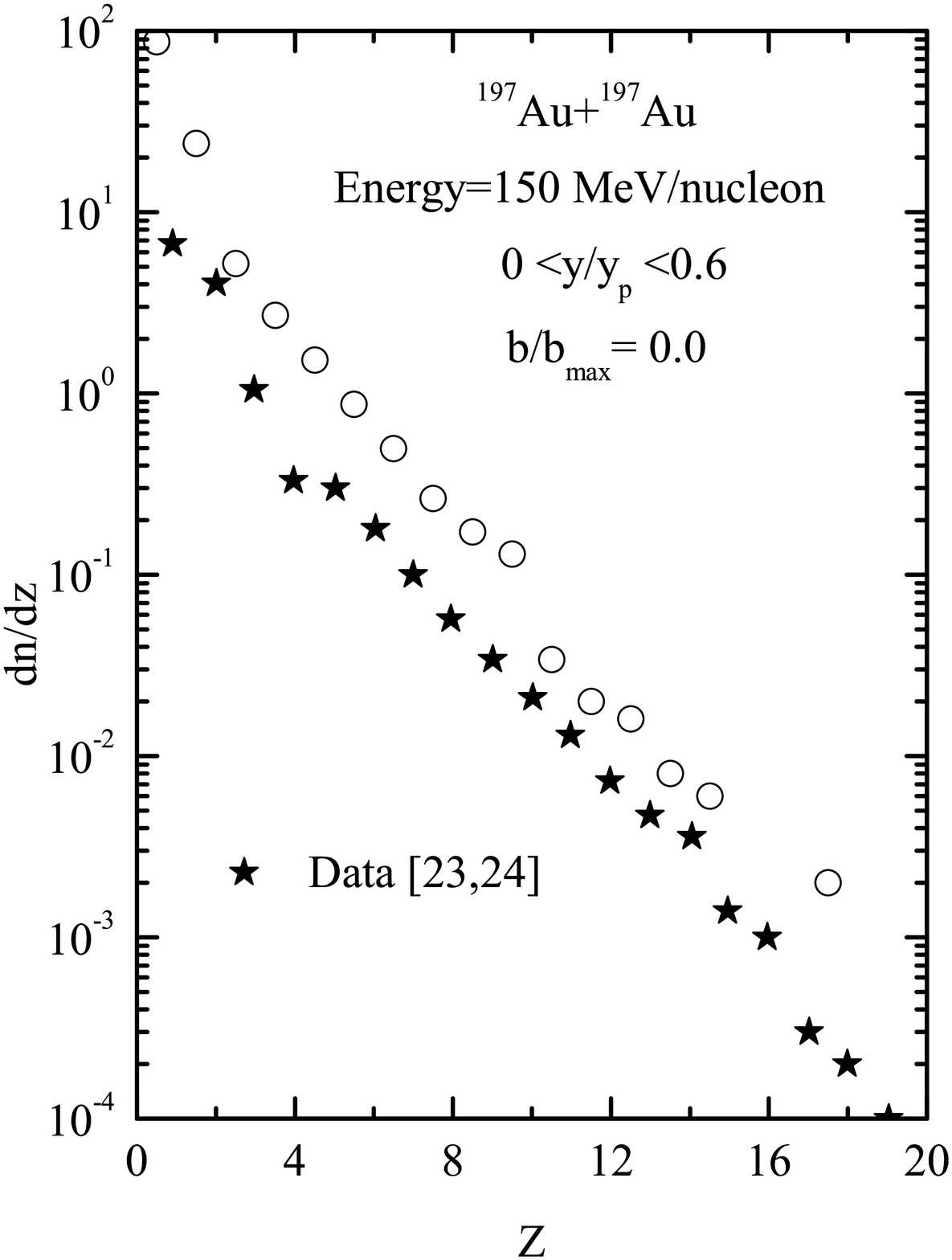}
 \vskip -0cm \caption{ The multiplicity distribution (dn/dz) of fragments measured for
  $^{197}Au$ +$^{197}$Au reaction at 150 MeV/nucleon having central
  collisions. The experimental data has been taken by the Ref. [23,24].}\label{fig3}
\end{figure}

\begin{figure}[!t]
\centering
 \vskip 1cm
\includegraphics[angle=0,width=12cm]{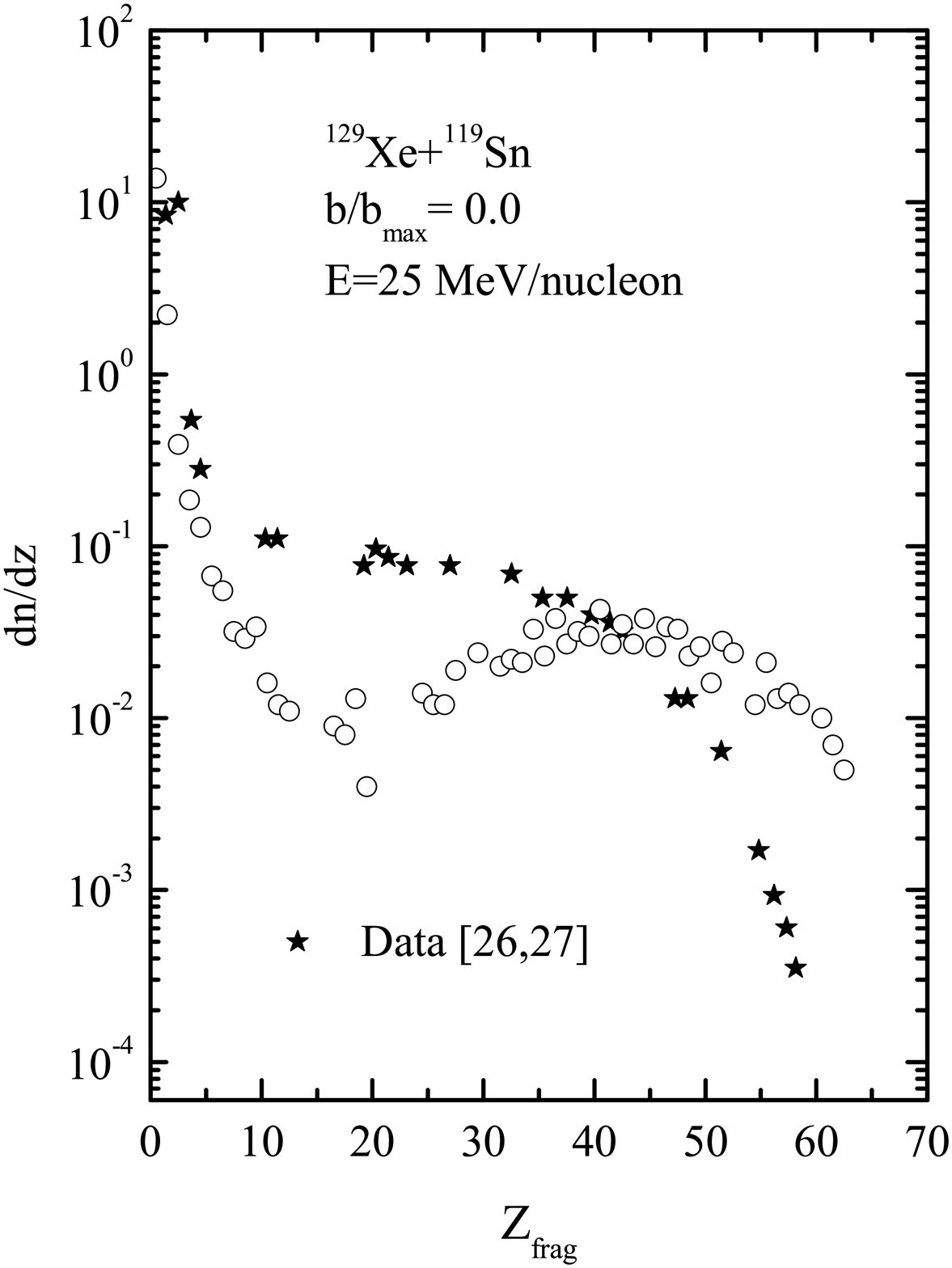}
 \vskip -0cm \caption{ The charge distribution for central collisions of
 $^{129}$Xe+$^{119}$Sn at bombarding energy of 25 MeV/nucleon. The experimental data has been taken by the Ref. [26,27].}\label{fig4}
\end{figure}

\begin{figure}[!t]
\centering
 \vskip 1cm
\includegraphics[angle=0,width=12cm]{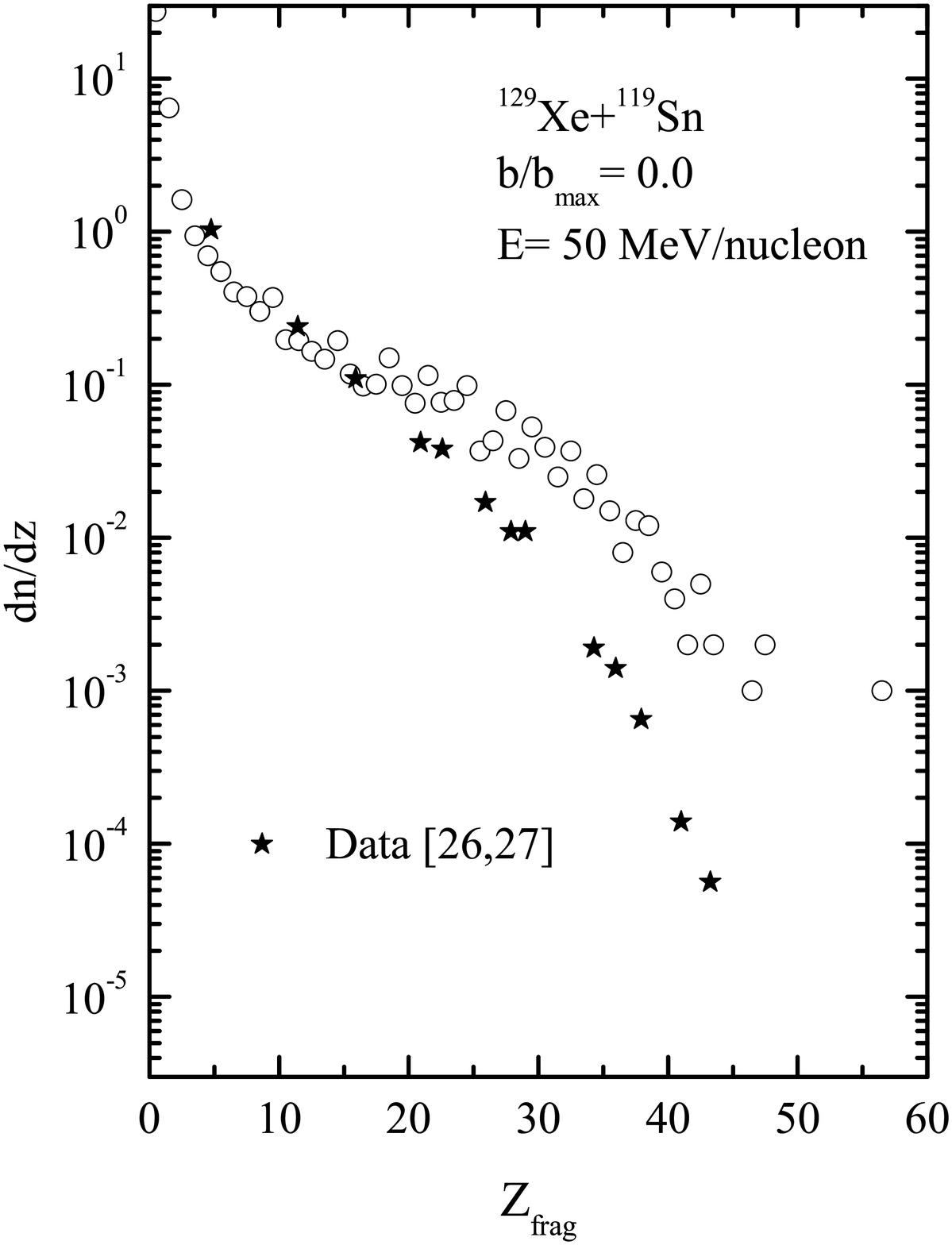}
 \vskip -0cm \caption{ Same as Figure 4, but at incident energy of E= 50 MeV/nucleon. The experimental data has been taken by the Ref. [26,27].}\label{fig5}
\end{figure}

\begin{figure}[!t]
\centering
 \vskip 1cm
\includegraphics[angle=0,width=12cm]{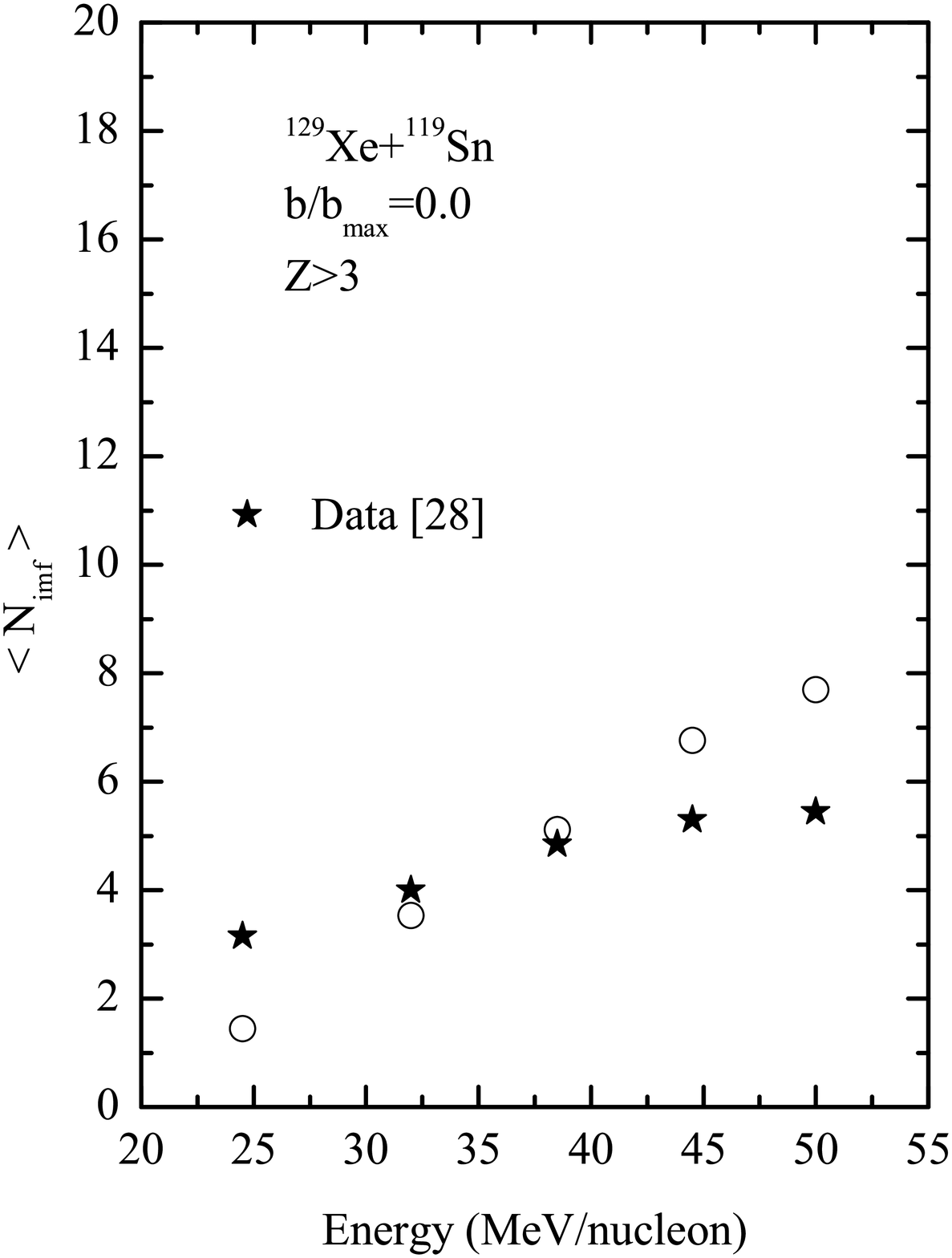}
 \vskip -0cm \caption{ The IMF multiplicity versus beam energy is shown for central collision in the reaction of $^{129}Xe$ +$^{119}$Sn. The experimental data has been taken by the Ref. [28].}\label{fig6}
\end{figure}
We simulate thousand events for the reactions of $^{40}Ar$
+$^{45}$Sc (at 85 and 115 MeV/nucleon), $^{197}Au$ +$^{197}$Au (
at 150 MeV/nucleon) and $^{129}$Xe+$^{119}$Sn (at 25, 50
MeV/nucleon). The impact parameters have been guided by the
corresponding experimental data. We use a soft equation of state
along with standard energy-dependent nucleon-nucleon Cugnon
cross-section. The details of the cross-section has been found in
Ref.   In fig. 1, we display the reaction of $^{40}Ar$ +$^{45}$Sc
at central collision at an incident energy of 85 MeV/nucleon.
Here, we display the charge distribution (open circles) for this
reaction. The figure shows a linear decrease in the value of
charge distribution (dn/dz) with the charge. This negative slope
of charge distribution indicates a gradual transition from the
spectator matter to the disassembly of the system.

\par
In the figure 2, we display the charge distribution for the
reaction of $^{40}Ar$ +$^{45}$Sc at central collision at an
incident energy of 115 MeV/nucleon. Here also, a decrease in the
value of charge distribution is observed. Also, when the beam
energy is increased from 85 MeV/nucleon to 115 MeV/nucleon, the
slope is still steeper indicates the total disassembly of the
matter. When we compare our theoretical results with experimental
data, we see that our theoretical calculations matches well with
the experimental data at both the energies. This experimental data
are taken with Michigan State University (MSU) FOPI array [20,21]
at the National Superconducting Cyclotron Laboratory (NSCL) using
beams from K1200 Cyclotron. The FOPI array consists of a main ball
of 170 phoswich counters covering angles from 23 degrees to 157
degrees.
\par
In fig. 3, we display the results for the central collisions of
$^{197}Au$ +$^{197}$Au collisions. This reaction was measured at
the incident energy of E=150 MeV/nucleon. Also, the condition of
rapidity cut is applied so that there may be the covering of only
those particles which are emitted from the most violent reactions.
This cut also removes the spectator particles i.e. only those
particles are taken which took part in the reaction.
\par
In the figure, we again see a decrease in the value of charge
distribution (dn/dz) with the increase in charge. The reason is
same as mentioned earlier in the description of figure 1. In the
figure, the open circles show our theoretical calculations. When
we compare our theoretical calculations with the experimental
data, it matches well with each other. These experimental data are
taken with the phase 1 set up of FOPI facility. The Au ion beam
was delivered by the rapid-cycling synchrotron SIS, Darmstadt
[22-24].
\par
In fig. 4, we display the central collisions (b/b$_{max}$=0.0) of
$^{129}Xe$ +$^{119}$Sn at bombarding energy of E=25 MeV/nucleon.
Open circles shows our theoretical calculations done by the MST
approach. In compare to previous results, we see now two separate
zones. First, a negative slope for light fragments and 'u' shape
for heavy fragments. This indicates that at low energy, reaction
produces excited compound system, thus cool down with the emission
of free nucleons. Nearly, no intermediate mass fragment is seen in
the reaction.
\par
In  figure 5, we again display the central collisions
(b/b$_{max}$=0.0) of $^{129}Xe$ +$^{119}$Sn at bombarding energy
of E= 50 MeV/nucleon. We again see a similar steepening of charge
distribution with the charge. When we compare our theoretical
calculations with the experimental data taken by the INDRA at the
GANIL and GSI accelerator [25-27], they are in good agreement with
each other.
\par
In figure 6, we  display the reaction of $^{129}$Xe +$^{119}$Sn
reaction for central collisions. We take only those particles with
Z>3. In the figure, we have seen the variation of number of
intermediate mass fragments with the increase in the beam energy.
At low energy i.e. 25 MeV/nucleon, number of IMFs produced is less
because we have mainly the heavy fragment and free nucleons. Also,
there is more time for the reaction to take place at low energy as
compared to high energy. But with the increase in energy, there is
increase in IMFs as shown in the figure. When our theoretical
calculations are matched with the experimental data taken during
the fragmentation studies performed with FOPI detector [28], they
are in good agreement.

{\Large \textbf{Summary}}\\
We studied the different symmetric and asymmetric reactions
 mainly for central collisions with different range of energies.
 We observed the steepening of charge distribution with the
 increase in charge. Also, we observed that number of IMFs
 increases with the increase in beam energy. Our calculations
 showed good agreement with the data at all the energies.

\end{document}